\documentclass[12pt]{article}
\usepackage[]{graphicx}\usepackage[]{color}
 
\usepackage{alltt}  
\usepackage{pifont} 
\usepackage{url}
\usepackage{breakurl}
\usepackage{xcolor}
\usepackage[colorlinks = true,
            linkcolor = blue,
            urlcolor  = red!80!blue,
            citecolor = blue!80!black,
            anchorcolor = blue]{hyperref}
\usepackage{soul}
\usepackage{paralist}
\usepackage{bm}
\usepackage[round]{natbib}
\usepackage{graphicx}
\usepackage{amsmath, wrapfig,amssymb,multirow}
\usepackage[margin=1in]{geometry}
\usepackage[parfill]{parskip}
\usepackage{epsfig, subfigure}
\usepackage{amsfonts}
\usepackage{authblk}

\usepackage{booktabs}
\usepackage{bigstrut}
\usepackage{threeparttable} 
\usepackage[format=hang,labelfont=bf]{caption}
\usepackage{float}
\floatstyle{boxed}
\usepackage{footnote}
\usepackage[inline]{enumitem}

\makesavenoteenv{tabular}
\makesavenoteenv{table}
\let\tempone\itemize
\let\temptwo\enditemize
\renewenvironment{itemize}{\tempone\addtolength{\itemsep}{0.5\baselineskip}}{\temptwo}
\IfFileExists{upquote.sty}{\usepackage{upquote}}{}

\usepackage[parfill]{parskip}

\begin{document}

\title{Quantifying the Contributions of Training Data and Algorithm Logic to the Performance of Automated Cause-assignment Algorithms for Verbal Autopsy}

\author[1,2,3]{Samuel J. Clark}
\author[4]{Zehang Li}
\author[4,5]{Tyler H. McCormick}

\affil[1]{Department of Sociology, The Ohio State University, Columbus, OH, USA}
\affil[2]{MRC-Wits Rural Public Health and Health Transitions Research Unit (Agincourt), School of Public Health, Faculty of Health Sciences, University of the Witwatersrand, Johannesburg, South Africa}
\affil[3]{INDEPTH Network, Accra, Ghana}
\affil[4]{Department of Statistics, University of Washington, Seattle, WA, USA}
\affil[5]{Department of Sociology, University of Washington, Seattle, WA, USA}

\maketitle

\begin{abstract}  
\noindent 
A verbal autopsy (VA) consists of a survey with a relative or close contact of a person who has recently died.  VA surveys are commonly used to infer likely causes of death for individuals when deaths happen outside of hospitals or healthcare facilities.  Several statistical and algorithmic methods are available to assign cause of death using VA surveys.  Each of these methods require as inputs some information about the joint distribution of symptoms and causes.  In this note, we examine the generalizability of this symptom-cause information by comparing different automated coding methods using various combinations of inputs and evaluation data.  VA algorithm performance is affected by both the specific SCI themselves and the logic of a given algorithm.  Using a variety of performance metrics for all existing VA algorithms, we demonstrate that in general the adequacy of the information about the joint distribution between symptoms and cause affects performance at least as much or more than algorithm logic.
\end{abstract}

\section{Introduction}

Verbal autopsy (VA) algorithms rely on three components:
\begin{enumerate*}[label=(\roman*)]
\item {VA data},
\item {symptom-cause information (SCI)}, and
\item {an algorithm or probabilistic method that combines the two to identify cause-specific mortality fractions (CSMF)s and/or assign a likely cause to each death}
\end{enumerate*}.  
The SCI describes how VA symptoms are related to each cause, i.e. the SCI provides information about the joint distribution of symptoms and causes.  SCI can be learned by using `labeled deaths' (deaths with both VA and a cause assigned through an independent mechanism) as training data in a typical statistical learning setting, or SCI can be a variety of expert-derived information.  For a Naive Bayes classifier, for example, SCI could consist of conditional probabilities of symptoms given a cause obtained by consensus from a group of physicians, $\Pr(s|c)$.

Predictive performance is usually used to compare VA algorithms, either at the individual level for cause of death assignment, or at the population level for estimating CSMFs.  When labeled deaths are available, there is a training set $\mathcal{T} = \{(x_1, y_1), ..., (x_n, y_n)\}$ where each $x$ denotes a symptom/indicator and each $y$ denotes a known cause of death.  In this situation a VA algorithm can be considered as a model $\hat f$ so that the predicted label for a new dataset $(X, Y)$ is  $\hat f(X)$. The performance of an algorithm conditional on the training data can then be denoted as 
\[
\mbox{Metric}_{\mathcal{T}} = E_{X, Y}[g(Y, \hat f(X^0)) | \mathcal{T}].
\]
This quantity refers to the conditional performance of an algorithm given a fixed training dataset, which is typically difficult to estimate directly with standard techniques such as cross-validation~\citep[see, e.g., discussions in][]{friedman2001elements}. In practice, training data in the VA context are very difficult to obtain, and researchers typically have to choose a single source of `gold-standard' data~\citep[e.g.,][]{king2008verbal,james2011performance,serina2015improving,miasnikof2015naive} or a specific set of expert opinions (e.g. the conditional probabilities mentioned above)~\citep{byass2012strengthening,mccormick2016probabilistic} to fit the algorithms. Therefore fair comparisons can only be achieved when the same SCI (training data or otherwise) is fixed for all algorithms in the comparison.


Using a large set of labeled VA deaths as training data, we empirically compare the performance of six variations of the common VA algorithms using different training and testing data. We demonstrate that the choice of training data plays \textit{the} crucial role in algorithm performance.

\section{Data}
The Population Health Metrics Research Consortium (PHMRC) dataset \citep{murray2011population} contains $7,841$ adult deaths that occurred in hospitals in six populations (Andhra Pradesh, India; Bohol, Philippines; Dar es Salaam, Tanzania; Mexico City, Mexico; Pemba Island, Tanzania; and Uttar Pradesh, India) in the years leading up to 2011, see Table~\ref{tab:sites}.  Each death has both VA data and medically-certified causes of death.  Each death has $251$ VA symptom items and $678$ stem word indicators extracted from the free text recorded by the interviewer. The medically-certified causes are provided in three levels of aggregation consisting of 55, 46, and 34 causes. We use the highest level cause list with 34 causes. The data cleaning procedures are described in~\citet{mccormick2016probabilisticSupplement}.

\begin{table}[ht]
\centering
\caption{PHMRC Deaths.}
\begin{tabular}{lr}
\toprule
Site & Deaths \\
\midrule
Andhra Pradesh & 1,554 \\
Bohol & 1,259 \\
Dar es Salaam & 1,726 \\
Mexico City & 1,586 \\
Pemba Island & 297 \\
Uttar Pradesh & 1,419 \\ 
\bottomrule
\end{tabular}
\label{tab:sites}
\end{table}

\section{Method}
In this section we describe our procedure for comparing the relative importance of the three components required for VA algorithms.  In particular, we leverage heterogeneity across the PHMRC sites to examine the sensitivity of VA algorithms to different SCIs.  We show that, while there is variation across VA algorithms, the crucial component driving performance of VA methods is the SCI. 

\subsection{Train-test split} \label{sec:train-test}
We generated training and testing data pairs in two ways. First we took all the observations from one site as the training set, and all the observations from another site as the testing set. We include here the case where we use the same site for both training and testing, resulting in $6\times 6$ tests. 
A possible concern is that the CSMFs can be dramatically different across different sites, and may be a main effect that drives the performance metrics. Thus in a second set of tests, we resample in each testing set so that the true CSMFs in the testing data are different and more diffuse in each replication. We first generated a distribution of causes from a $\mbox{Dirichlet}(1)$ distribution and then resampled with replacement within testing sets so that the resampled datasets match the sampled distribution of causes. We repeated this resampling step of testing data $50$ times for each train-test split to obtain the average performance metrics over the $50$ replications, leading also to $6\times 6$ tests. 

\subsection{VA algorithms}
We fit five commonly used algorithms to each training and testing spit.  The algorithms are:
\begin{enumerate}
   \item Tariff 1.0 (\cite{james2011performance}) implemented based on code in our open-source replication \texttt{R} package \citep{tariffCRAN2018}.
   \item InterVA  (\cite{byass2012strengthening,Fottrell2007}) with the physician-provided conditional probabilities, $\Pr(s|c)$, replaced by conditional probabilities recalculated from the training data, and then turned into ranks by choosing cutoff values so that the quantiles of each level match that in the original software. We denote this method as InterVA-Q.  InterVA-4 is implemented based on code in our open-source \texttt{R} package \citep{interVACRAN2018}.
   \item  InterVA with the physician-provided conditional probabilities, $\Pr(s|c)$, replaced by conditional probabilities recalculated from the training data, and then turned into ranks by finding the closest level with fixed value interpretations as in the original software.  We denote this method as InterVA-F.  InterVA-F is implemented based on code in our open-source \texttt{R} package \citep{interVACRAN2018}.
   \item InSilicoVA  (\cite{mccormick2016probabilistic,mccormick2016probabilisticSupplement}), using the same SCI  as InterVA-Q. We denote this method as InSilicoVA-Q.    InSilicoVA-Q is implemented based on code in our open-source \texttt{R} package `InSilicoVA' \citep{inSilicoVACRAN2018}.
   \item InSilicoVA, using the same SCI as InterVA-F. We denote this method as InSilicoVA-F.  InSilicoVA-F is implemented based on code in our open-source \texttt{R} package `InSilicoVA' \citep{inSilicoVACRAN2018}
 \end{enumerate} 
{}
\subsection{Performance metrics}
We compare the performance of VA algorithms using four metrics. 
\begin{itemize}
  \item {\bf Overall chance-corrected concordance (CCC)}. CCC for cause $j$ is defined as
  \[ CCC_j = \frac{\frac{TP_j}{TP_j + TN_j} - \frac{1}{C}}{1 - \frac{1}{C}}
  \] 
  where $TP_j$ is the number of true positives for cause $j$, and $TN_j$ is the number of true negatives for cause $j$. It is worth noting that the definition of $TN_j$ is the the number of cases where the cause assigned to a death is not cause $j$ while the true cause is cause $j$.   So  CCC could also be written as
  \[ CCC_j = \frac{\frac{\# \text{ correctly assigned to cause }  j}{\# \text{ total number of death from cause } j } - \frac{1}{C}}{1 - \frac{1}{C}}.
  \] 
  Then the overall CCC is defined as a weighted sum of cause-specific CCC. Three ways to construct the weight is discussed in \citet{murray2011robust}, and we follow the recommendation and used equal weights is this study. 

  \item {\bf CSMF accuracy}
    \[ ACC_{csmf} = 1 - \frac{\sum_{c=1}^C |CSMF^{true}_c - CSMF^{pred}_c | }{2(1 - \min CSMF^{true})}
    \]
    This form was defined in \citet{murray2011robust}. The idea is that the worst possible case for CSMF prediction is to put all the weight on the minimum CSMF value that corresponds to a total absolute error of $2(1 - \min CSMF^{true})$. So $ACC_{csmf}$ has a value between 0 and 1. 

  \item {\bf Top cause accuracy}
    \[ ACC_1 = \frac{\text{\# of correct COD being first cause assignment}}{N}
    \]
  \item {\bf Top 3 cause accuracy}
    \[ ACC_3 = \frac{\text{\# of correct COD within first three cause assignments}}{N}
    \]
  
\end{itemize}

\section{Results}
\begin{figure}
\includegraphics[width=\textwidth]{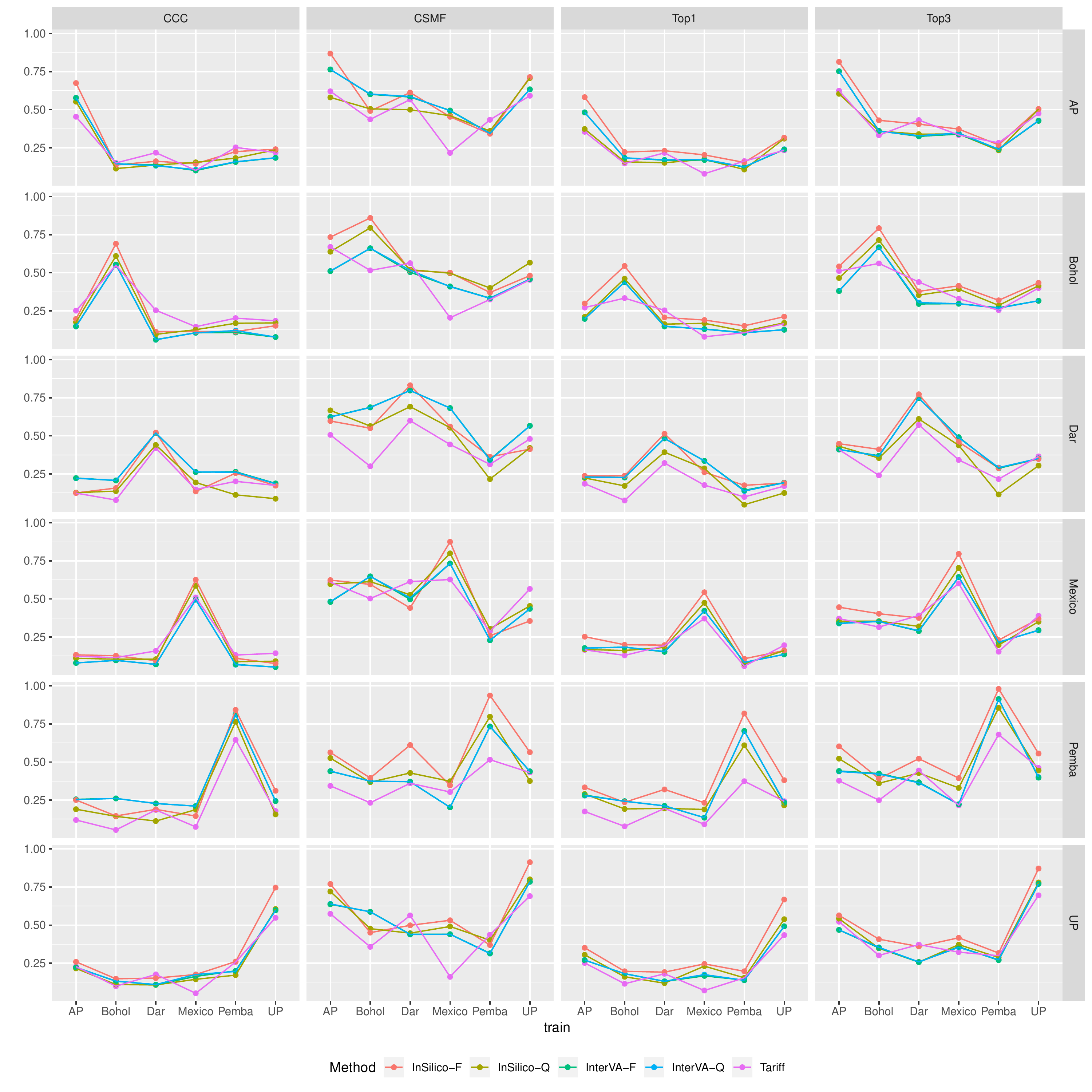}
\caption{VA algorithm performance for each train-test split defined in the first (non resampled) experiment design. Test site on the vertical axis, training site on the horizontal axis, one column for each performance metric, and different colors  for each algorithm.}
\label{fig:measure1}
\end{figure}

\begin{figure}
\includegraphics[width=\textwidth]{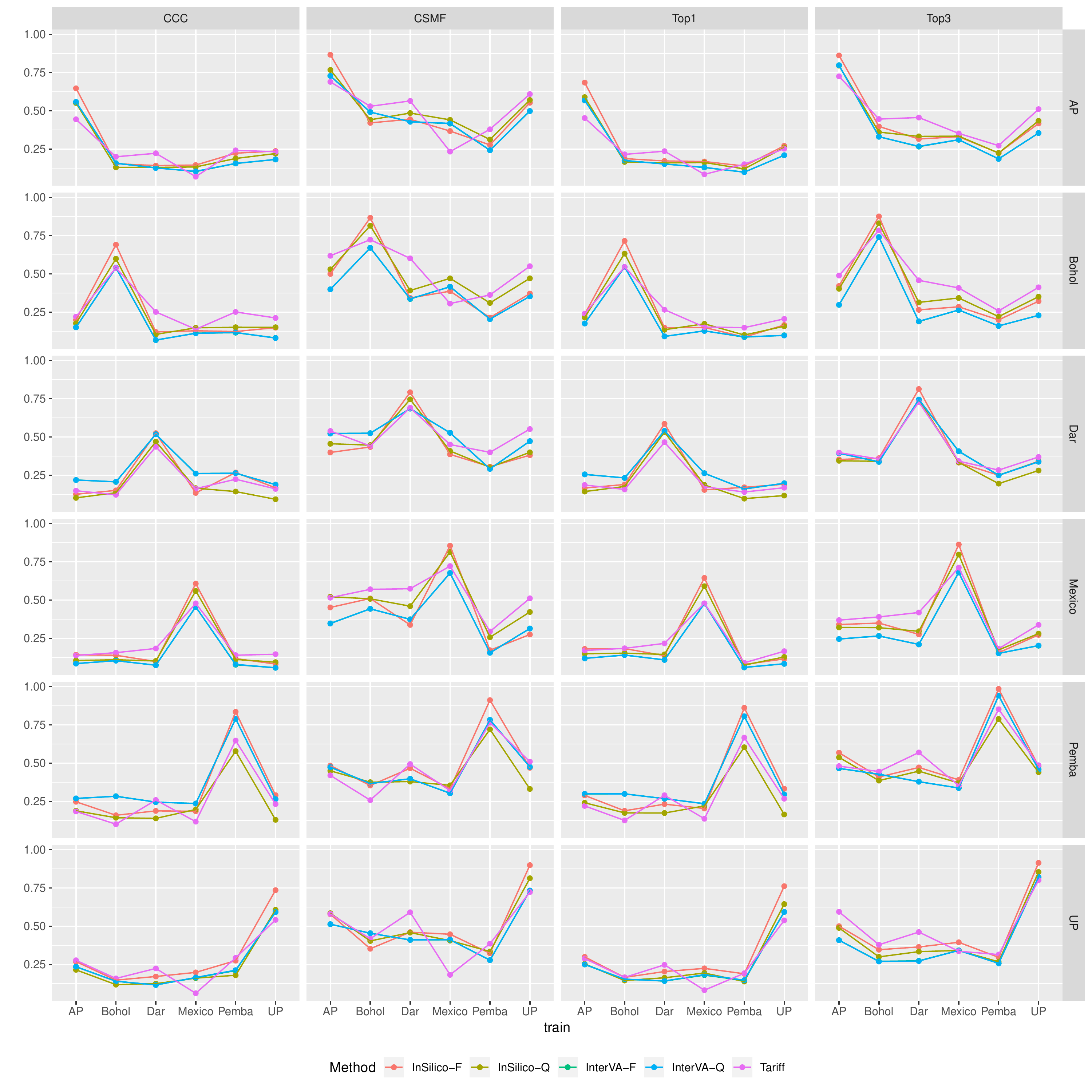}
\caption{Average VA algorithm performance for each train-test split defined in the second (resampled) experiment design. Data are resampled with replacement in each test site to create different CSMFs in each replication. Test site on the vertical axis, training site on the horizontal axis, one column for each performance metric, and different colors  for each algorithm.}
\label{fig:measure2}
\end{figure}

VA algorithm performance for both experiment designs described in Section \ref{sec:train-test} is summarized in Figures~\ref{fig:measure1} and~\ref{fig:measure2}.  Though there is some variation in which method performs the best across metrics and train/test combinations, one striking pattern is that the performance for all methods is best when trained and tested on the same site.  This pattern remains after performing the resampling procedure we described, indicating that the difference is related to the joint distribution of symptoms and causes and not to discrepancies only in the marginal cause distributions across sites.

 \section{Proportion of the explained variance}
 In this section we address the question: \textbf{how much of the overall variation in performance metrics is associated with the training data compared to algorithm logic?} We used analysis of variance (ANOVA) on each of the four metrics. The total variation in each metric is divided into the variation between the algorithms, training data, testing data, and the residual (error). Next we describe the detailed model and results. 

 \subsection{Testing all sites}
First we performed the joint analysis of results obtained from all the testing sites. In the first set of experiments there are 180 ($6 \mbox{ training site } \times 6 \mbox{ testing site } \times 5 \mbox{ algorithm}$) scores for each performance measure. When trained and tested on the same dataset (or same dataset after resampling), we expect that the within-sample performance will be better, and thus we further adjusted for such possible exaggeration of performance with an additional additive term and compared with performance after removing all self-training results.

We used a simple additive model to explore the proportion of variation explained by training data, testing data, and the choice of algorithm:  
\[
  Y_{ijk} = \mu + \alpha_i + \beta_j + \delta_k + \gamma \bm{1}_{i=j} + \epsilon_{ijk}, 
\] 
where $Y_{ijk}$ is the performance measurement for algorithm $k$ trained on site $i$ and tested on site $j$, and one level in each of the $\alpha_i, \beta_j$, and $\delta_k$ terms is chosen as the baseline. We decompose $Y_{ijk}$ into a baseline ($\mu$), a term specific for each site when used as training ($\alpha_i$), each site when used as testing site ($\beta_j$), the algorithm ($\delta_k$), an effect, $\gamma$, for when the testing and training site are the same ($\bm{1}_{i=j}$), and residual error ($\epsilon_{ijk}$). In a set of four experiments we fit the simple linear model to:
\begin{enumerate}
    \item The first set of tests without resampling including all $36$ train-test pairs.
    \item The second set of resampled tests with average measurement over $50$ replications including all $36$ train-test pairs.
    \item The first set of tests without resampling excluding the measurements where algorithms are trained and tested on the same dataset.  For this we remove $\gamma$ from the model as well as any results where the training and testing site are the same.
    \item The second set of resampled tests with average measurement over $50$ replications excluding the measurements where algorithms are trained and tested on the same dataset.  For this we remove $\gamma$ from the model as well as any results where the training and testing site are the same.
  \end{enumerate}  

\begin{figure}
\centering
\includegraphics[width=\textwidth]{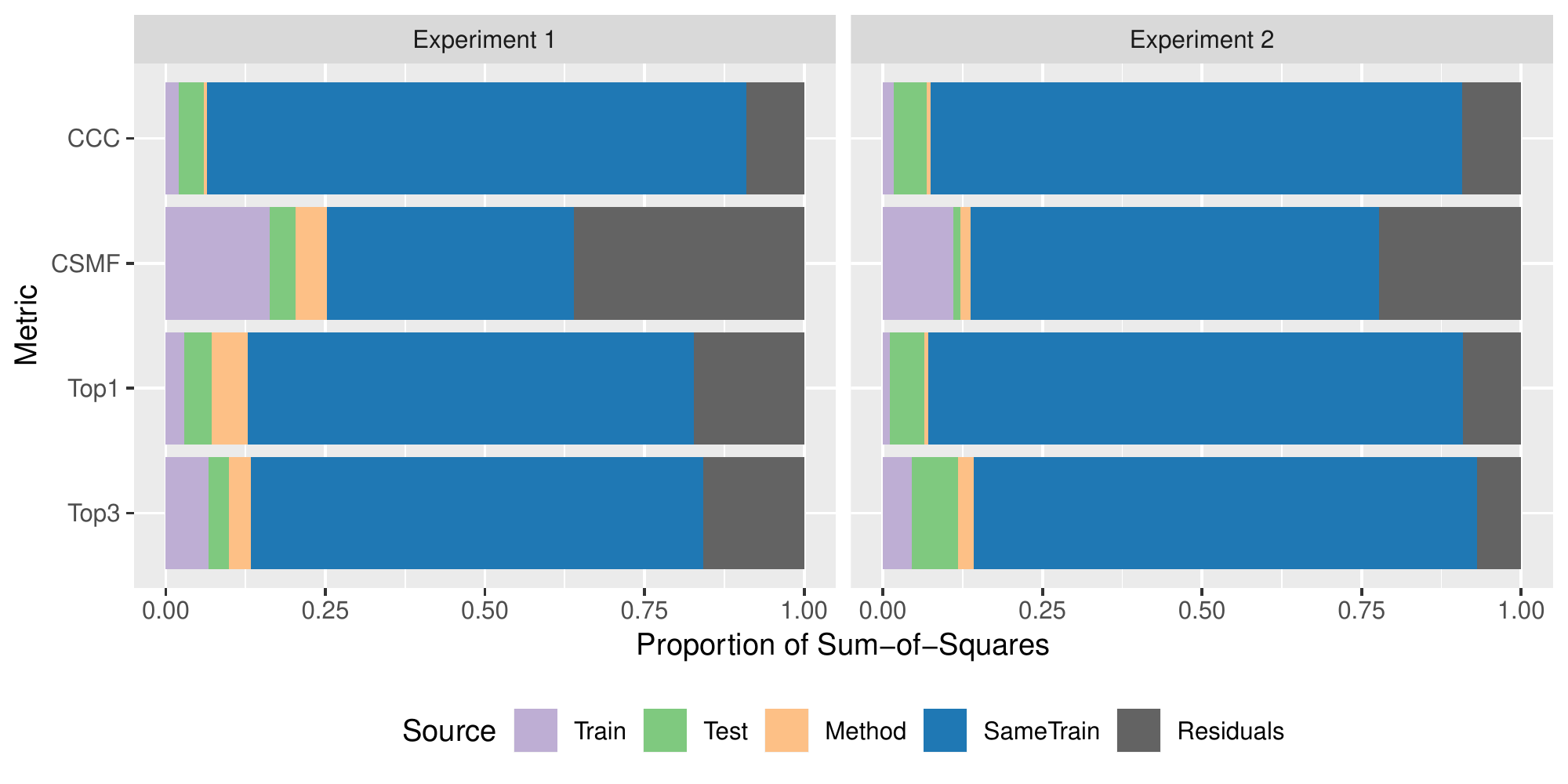}
\includegraphics[width=\textwidth]{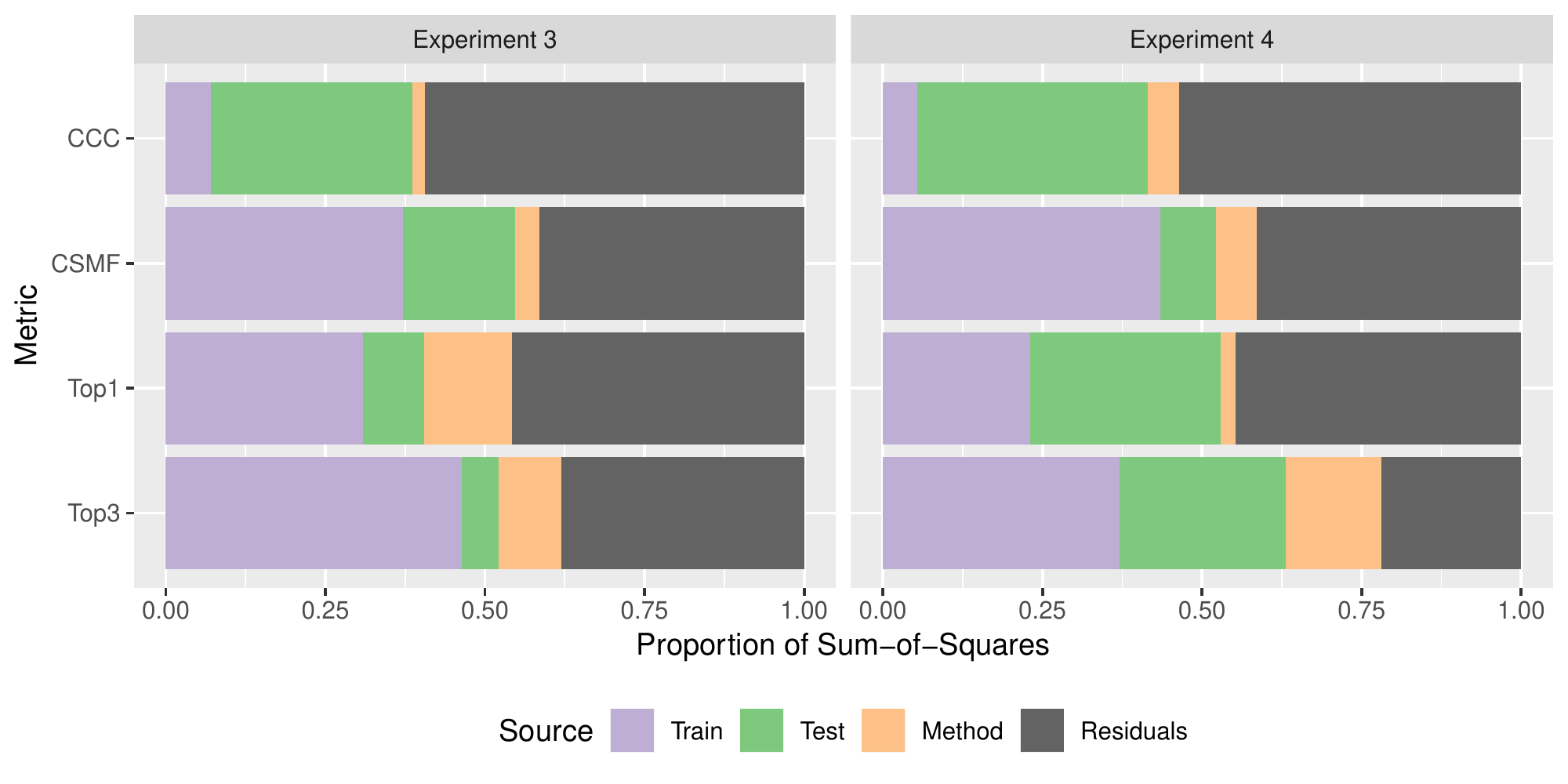}
\caption{Decomposition of variance in the four experiments.  Experiment 1 includes tests without resampling including all $36$ train-test pairs.  Experiment 2 uses resampled tests with average measurement over $50$ replications including all $36$ train-test pairs.
  Experiment 3 includes tests without resampling excluding the measurements where algorithms are trained and tested on the same dataset.  For experiment 4 we use resampled tests with average measurement over $50$ replications excluding the measurements where algorithms are trained and tested on the same dataset.}
\label{fig:var}
\end{figure}

In all the four experiments we are interested in the proportion of variation explained by coefficient $\alpha_i$, i.e. the choice of training data. In all four experiments $p$-values associated with $\alpha_i$ are mostly tiny, as shown in Table~\ref{tab:pvalue1}. The decomposition of total variation is summarized in Figure~\ref{fig:var}. In experiments 1 and 2, whether or not the training and testing data are from the same site explains the overwhelming fraction of the total variation, which is not surprising given the results in Figures~\ref{fig:measure1} and~\ref{fig:measure2}. When the same-site train/test splits are removed, it is clear that the choice of training data still explains a large
proportion of overall variation, see the lower panels of Figure~\ref{fig:var}.  


For all metrics except CCC, in the experiments without same-site train/test splits, the training data explain a significantly larger fraction of overall variation compared to the algorithms in experiment 3 without resampling.  In experiment 4 with resampling of the testing data, the fraction of variation explained by training data and algorithm logic is at least similar for these three metrics, and again in some cases, more variation is explained by the training data.

\begin{table}[ht]
\centering
\captionsetup{margin=3.75cm}
\caption{Estimated $p$-values for the effect of training site, $\alpha_i$, in the four experiments for each of the four performance metrics.}
\begin{tabular}{ccccc}
  \toprule
 Experiment & CCC & CSMF & Top1 & Top3 \\ 
  \midrule
 1 & 0.0000 & 0.0000 & 0.0001 & 0.0000 \\ 
   2 & 0.0000 & 0.0000 & 0.0023 & 0.0000 \\ 
   3 & 0.0091 & 0.0000 & 0.0000 & 0.0000 \\ 
   4 & 0.0215 & 0.0000 & 0.0000 & 0.0000 \\ 
   \bottomrule
\end{tabular}
\label{tab:pvalue1}
\end{table}

\subsection{Testing individual sites}\label{sec:anova-sites}
\begin{figure}
\centering
\includegraphics[width=\textwidth]{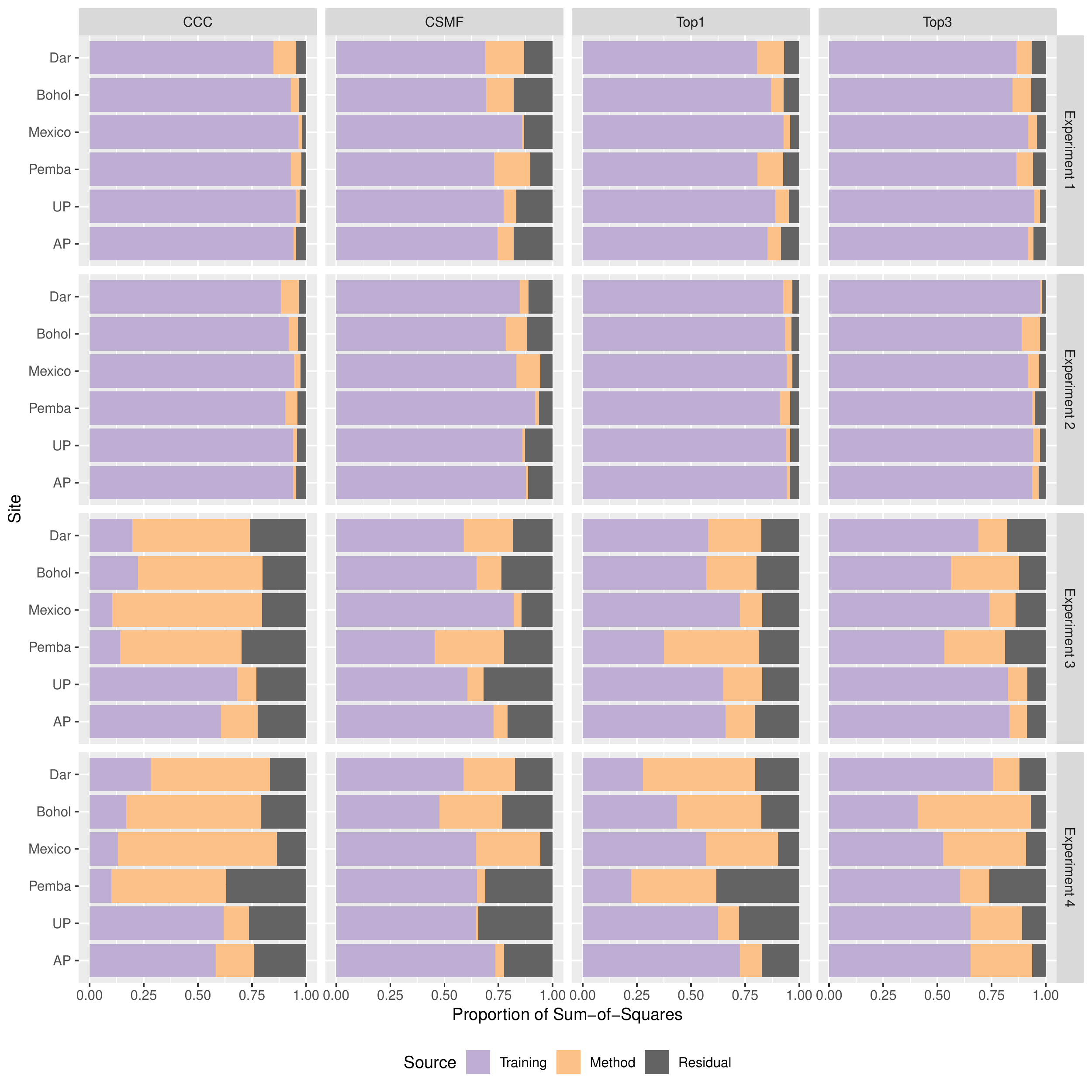}
\caption{Decomposition of variance in experiments 1 to 4 \textit{fitted to data from each testing site}. 
Experiment 1 includes tests without resampling including all measurements tested on each site.  Experiment 2 uses resampled tests with average measurement over $50$ replications tested on each site. 
Experiment 3 includes tests without resampling excluding the measurements where algorithms are trained and tested on the same dataset.  For experiment 4 we use resampled tests with average measurement over $50$ replications excluding the measurements where algorithms are trained and tested on the same dataset.}
\label{fig:var2}
\end{figure}

To further remove any nonlinear effects from the test set, we also examined the same variation decomposition when fitting the linear model to any single site as test data, i.e. we fit the model
\[
  Y_{ijk} = \mu + \alpha_i + \delta_k + \epsilon_{ik}, 
\]
for each test site $j = 1, ..., 6$ separately. In this case, $\beta_j$ and $\gamma$ are dropped from the model because of collinearity. In the models for each site, Experiments 1 and 2 reduce to the same form as Experiments 3 and 4, with the only difference being whether or not the same-site training data are included. We repeated the same Experiments as in the previous section. This decomposition of variation is summarized in Figure~\ref{fig:var2}. Similar to before, when results from same-site train/test splits are included, the choice of training data explains most of the variation in performance (experiments 1 and 2). After removing the results from same-site train/test splits (experiments 3 and 4), the proportion of variation explained by the choice of training data and the choice of algorithm varies, with training data generally being more important.

\begin{figure}
\centering
\includegraphics[width=\textwidth]{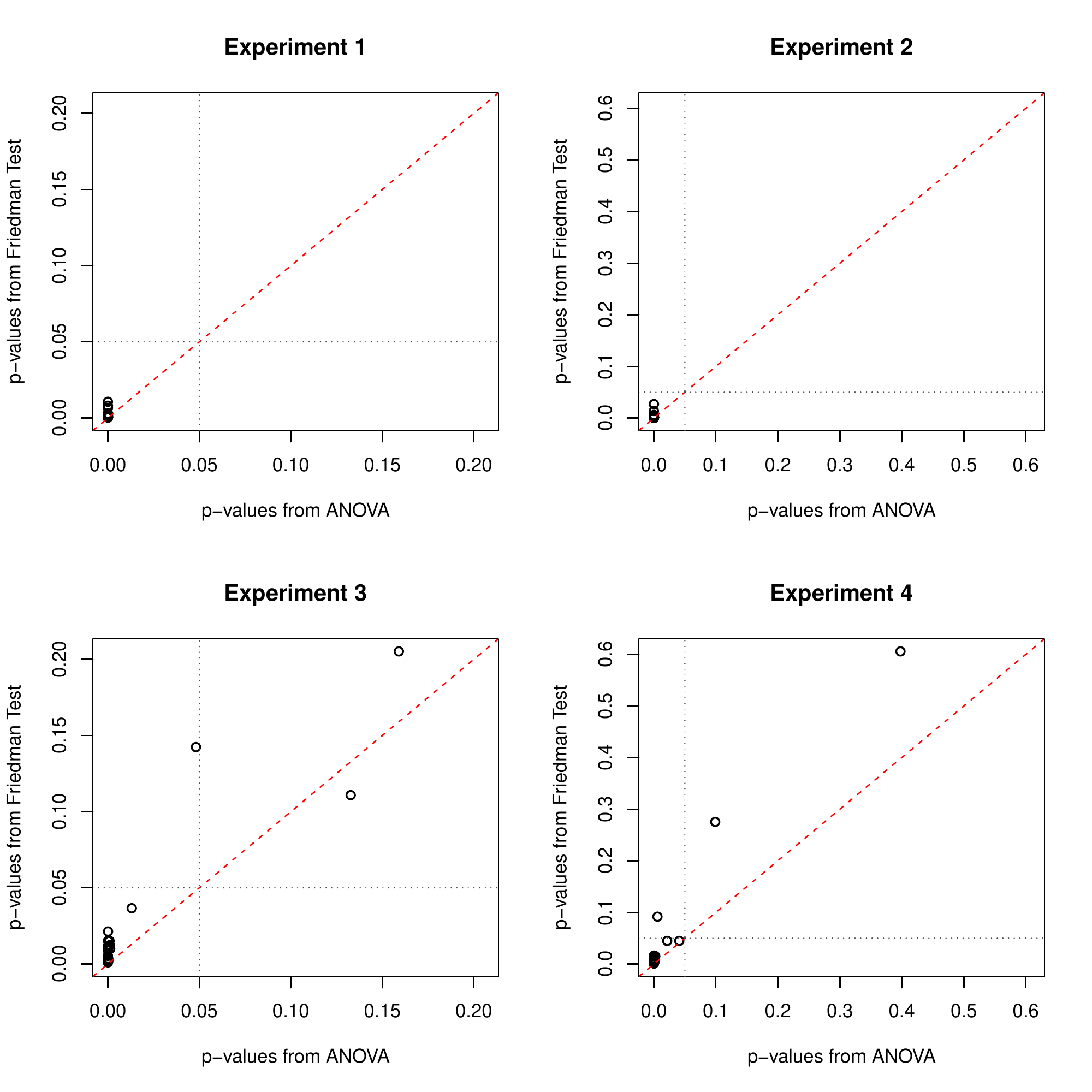}
\caption{Comparing the $p$-values associated with $\alpha_i$ in experiments 1 to 4 obtained from ANOVA and Friedman test.}
\label{fig:friedman}
\end{figure}

\section{Discussion}
The empirical analysis using the PHMRC dataset of labeled VA deaths clearly illustrates that the performance of VA algorithms depends strongly on the SCI.

The use of ANOVA may suffer from the potential violation of the normality and equal variance assumptions. However, for the purpose of illustration the variance decomposition still demonstrates the significant variation due to the choice of SCI. For a more powerful test, one might instead use the Friedman test~\citep{friedman1937use}, which is the non-parametric procedure similar to ANOVA. We also compared the $p$-values obtained from the models in Section~\ref{sec:anova-sites} to the $p$-values reported by the Friedman test and found only a few changes in the site-specific conclusions when results from same-site train/test splits are removed (i.e., Experiment 3 and 4), as illustrated Figure~\ref{fig:friedman}. 

We have demonstrated that the resampling of the cause of death distribution within testing data made little difference in terms of the large variation associated with the SCI.
However since the sample size of each cause within one site can be small, it is unclear how much impact a few rare causes can have on the performance metrics. Further experiments with a reduced set of major causes may also be beneficial to further understand the variation decomposition. 

It is clear from our results that all algorithms performed far better when trained and tested on the same site, and that even when same-site training data were not used, the choice of SCI was at least as important as the choice of algorithm.  This result leads to a strong recommendation: \textbf{the choice of SCI is critical to the performance of all VA algorithms, and all VA algorithms perform far better when trained on deaths from the same population.}  This should lead VA practitioners to prioritize the creation and maintenance of an SCI repository containing deaths from a wide variety of settings that is kept up to date and includes deaths from each population for which VA is used to assign causes of death.


\bibliographystyle{chicago}
\bibliography{lancetVA}

\end{document}